\documentclass[gmdd, hvmath, online,fleqn]{copernicus_discussions}%
\usepackage{hyperref}
\usepackage{graphicx}
\usepackage{amsmath}
\usepackage{amssymb}
\begin{document}
\title{Coupled atmosphere-wildland fire modeling with WRF-Fire
\\ version 3.3}
\author[1]{Jan Mandel}
\author[1]{Jonathan D. Beezley}
\author[2]{Adam K. Kochanski}
\affil
[1]{Department of Mathematical and Statistical Sciences, University of Colorado Denver, Denver, CO, USA}
\affil
[2]{Department of Meteorology, University of Utah, Salt Lake City, UT, USA}
\runningtitle{Coupled atmosphere-wildland model WRF-Fire 3.3}
\runningauthor{J. Mandel, J. D. Beezley, and A. K. Kochanski}
\correspondence{Jan Mandel\\ (jan.mandel@gmail.com)}
\received{}
\pubdiscuss{}
\revised{}
\accepted{}
\published{}
\firstpage{1}
\maketitle

\begin{abstract}\vspace*{-1.3mm}
We describe the physical model, numerical algorithms, and software structure
of \mbox{WRF-Fire}. \mbox{WRF-Fire} consists of a fire-spread model,
implemented by the level-set method, coupled with the Weather Research and
Forecasting model. In every time step, the fire model inputs the surface
wind, which drives the fire, and outputs the heat flux from the fire into the
atmosphere, which in turn influences the atmosphere. The level-set method allows
submesh representation of the burning region and flexible implementation of
various kinds of ignition. \mbox{WRF-Fire} is distributed as a part of WRF and
it uses the WRF parallel infrastructure for parallel computing.
\end{abstract}

\vspace*{-4.5mm}

\introduction\vspace*{-1.3mm} Wildland fires impact the lives of millions of
people and cause major damage every year worldwide, yet they are a natural
part of the cycle of nature. Better tools for modeling wildland fire behavior
are important for managing fire suppression, planning controlled burns to
reduce the fuels, as well as to help assess fire danger. Fire models range
from tools based on \citet{Rothermel-1972-MMP} fire spread rate formulas,
such as BehavePlus \citep{Andrews-2007-BFM} and FARSITE
\citep{Finney-1998-FFA}, suitable for operational forecasting, to
sophisticated \mbox{3-D} computational fluid dynamics and combustion
simulations suitable for research and reanalysis, such as FIRETEC
\citep{Linn-2002-SWB} and WFDS \citep{Mell-2007-PAM}. BehavePlus, the
PC-based successor of the calculator-based BEHAVE, determines the fire spread
rate at a single point from fuel and environmental data; FARSITE uses the
fire spread rate to provide a \mbox{2-D} simulation on a PC; while FIRETEC
and WFDS require a parallel supercomputer and run much slower than real time.

Wildland fire is a complicated multiscale process, from the flame reaction
zone on milimeter scale to the synoptic weather scale of hundreds of
kilometers. Since direct numerical simulation of wildland fire is
computationally intractable and detailed data are not available anyway,
compromises in the choice of processes to be modeled, approximations, and
parametrizations are essential. Fortunately, a practically important range of
wildland fire behavior can be captured by the coupling of a mesoscale
weather  model with a simple \mbox{2-D} fire spread model
\citep{Clark-1996-CAF,Clark-1996-CAM}. Weather has a major influence on
wildfire behavior; in particular, wind plays a dominant role in the fire
spread. Conversely, the fire influences the atmosphere through the heat and
vapor fluxes from burning hydrocarbons and evaporation of fuel moisture. Fire
heat output has a major effect on the atmosphere; the buoyancy created by the
heat from the fire can cause tornadic strength winds, and the air motion and
moisture from the fire can affect the atmosphere also away from the fire. It
is well known that a large fire ``creates its own weather''. The correct
wildland fire shape and progress result from the two-way interaction between
the fire and the atmosphere
\citep{Clark-1996-CAF,Clark-1996-CAM,Clark-2004-DCA,Coen-2005-SBE}.

WRF-Fire \citep{Mandel-2009-DAW} combines the Weather Research and
Forecasting Model (WRF) with the ARW dynamical core \citep{Skamarock-2008-DAR} with a semi-empirical fire
spread model. It is intended to be faster than real time in order to deliver
a prediction.

WRF-Fire has grown out of the NCAR's CAWFE code
\citep{Clark-1996-CAF,Clark-1996-CAM,Clark-2004-DCA,Coen-2005-SBE}. CAWFE
consists of the Clark-Hall mesoscale atmospheric model, coupled with a
tracer-based fire spread model. Although the Clark-Hall model has many good
properties, it is a legacy serial code, not supported, and difficult to
modify or use with real data, while WRF is a parallel supported community
code routinely used for real runs. See \citet{Coen-IWF} for a further
discussion of their relative merits in the wildland fire application.
WRF-Fire was started by \citet{Patton-2004-WCA}, who proposed a combination
of WRF with the tracer-based model from CAWFE, formulated a~road map, and
made the important observation that the innermost domain of the weather code,
which interacts directly with the fire model, needs to run in the Large Eddy
Simulation (LES) mode. Patton ported the Fortran 77-based fire module to
Fortran 90 and developed the initial serial coupled code. However, instead of
using the existing tracer-based CAWFE code, the fire module in
\mbox{WRF-Fire} was developed based on the level-set method
\citep{Osher-2003-LSM}. One of the reasons was that the representation of the
fire region by the level-set function was thought to be more flexible than
the representation of the burning region in CAWFE by four tracers in each
cell of the fire mesh. In particular, the level-set function can be
manipulated more easily than tracers for the purpose of data assimilation.
Insertion of the heat fluxes, while fundamentally the same as in CAWFE, had
to be redone for WRF variables already in Patton's initial code. Thus,
only the code for the calculation of the fire spread rate and the heat fluxes remained from
CAWFE. While \mbox{WRF-Fire} takes advantage of the experience accumulated
with CAWFE, WRF is quite different from the Clark-Hall atmospheric model and
the fireline propagation algorithm is also different. Thus, it needs to be
demonstrated that \mbox{WRF-Fire} can deliver similar results as CAWFE, and
WRF-Fire needs to be validated against real fires
(Sect.~\ref{sec:simulations}).

The level-set method was used for a surface fire spread model in
\citet{Mallet-2009-MWF}. \citet{Filippi-2009-CAF} coupled the atmospheric
model Meso-nh with fire propagation by tracers. Tiger \citep{Mazzoleni-T2F}
uses a \mbox{2-D} combusion model based on reaction-convection-diffusion
equations and a convection model to emulate the effect of the fire on the
wind. FIRESTAR \citep{Morvan-2004-MPW} is a physically accurate wildland fire
model in two dimensions, one horizontal and one vertical. UU LES-Fire
\citep{Sun-2009-IFC} couples the University of Utah's Large Eddy Simulation
code with the tracer-based code from CAWFE. See the survey by
\citet{Sullivan-2009-RWF} for a number of other models.

WRF-Fire was briefly treated as one of the topics in \citet{Mandel-2009-DAW}.
The purpose of this paper is to describe the fire module and the coupling
with WRF in the current \mbox{WRF-Fire} code in sufficient detail, yet
understandable to a reader not familiar with WRF. In addition, the advances
since the paper \citet{Mandel-2009-DAW} was written in 2007 include new,
practically important ignition schemes (Sect.~\ref{sec:ignition}), vertical
interpolation of the wind in the boundary layer dependent on land-use
(Sect.~\ref{sec:coupling}), parallel computing (Sect.~\ref{sec:software}),
input of real data (Sect.~\ref{sec:data-input}), and validations in progress
on real fires (Sect.~\ref{sec:simulations}). This paper also contains
reproducible descriptions of the physical model
(Sect.~\ref{sec:physical-model}), the required WRF settings
(Sect.~\ref{sec:WRF}), and the coupling of WRF with the fire module
(Sect.~\ref{sec:coupling}).

WRF-Fire is public domain software and it has been distributed as a part of the WRF source code at
\url{wrf-model.org} since version 3.2, released in April 2010
\citep{Dudhia-2010-WRF}. The released version is updated periodically and supported by NCAR.
The current development version of \mbox{WRF-Fire} with the latest features and bug fixes
and additional visualization tools, guides, and diagnostic utilities, are
available directly from the developers at \url{openwfm.org}. This article
describes \mbox{WRF-Fire} as it is scheduled to be included in WRF~3.3, to be released in
March or April 2011. \mbox{WRF-Fire} user's guide is available as a part of the WRF
user's guide \citep{Wang-2010-AUG}, to be updated with the release.

New features since WRF version 3.2 include new ignition models, vertical
interpolation of the wind from logarithmic profile, fetching high-resolution
geogrid data, terrain gradient interpolation, and optional input of fuel map,
land use map, and high-resolution topography in ideal runs. \vspace*{-3.5mm}

\section{Physical fire model and fuels \label{sec:physical-model}}

The physical model consists of functions specifying the fire spread rate and
the heat fluxes, and it is essentially the same as a subset of CAWFE
\citep{Clark-2004-DCA,Coen-2005-SBE}. The spread rate calculation is in turn based on BEHAVE
\citep{Rothermel-1972-MMP,Andrews-2007-BFM}. It is described here in more
detail for the sake of reproducibility and to point out the (minor)
differences.

\subsection{Fuel properties}

Fuel is characterized by the quantities listed in Table~\ref{tab:fuels},
which are given at every point of the fire mesh. To simplify the
specification of fuel properties, fuels are given as one of 13
\citet{Anderson-1982-ADF} categories, which are preset vectors of values of
the fuel properties. These values are specified in an input text file
(\texttt{namelist.fire}), and they can be modified by the user. The user can
also specify completely new, custom fuel categories.

\subsection{Fire spread rate}

The fire model is posed in the horizontal ($x$, $y$) plane the Earth surface
is projected on. The semi-empirical approach to fire propagation used here
assumes that the fire spread rate is given by the modified
\citet{Rothermel-1972-MMP} formula
\begin{equation} \label{eq:spread}
S=R_0\left( 1+\phi_{\rm W}+\phi_{\rm S}\right) ,
\end{equation}
where $R_0$ is the spread rate in the absence of wind, $\phi_{\rm W}$ is the
wind factor, and $\phi_{\rm S}$ is the slope factor. The components of
(\ref{eq:spread}) computed from the fuel properties (Table~\ref{tab:fuels}),
the wind speed $U$, and the terrain slope $\tan\phi$ following the equations
in Table~\ref{tab:spread}. See \citet{Rothermel-1972-MMP} for further
details, derivation, and justifications.

Chaparral is a special fuel in that the spread rate depends only on wind
speed. For chaparral, (\ref{eq:spread}) is replaced by
(\citealp{Coen-2001-CAM}, Eq.~1)
\begin{equation} \label{eq:chapparal}
S=\min\left\{ 6,\max\left\{ 0.03333, 1.2974U\right\} \right\} .
\end{equation}

The only differences from \citet{Rothermel-1972-MMP} are the subtraction of
the moisture from the fuel load in the computation rather than up front,
limiting the slope and the windspeed, the special chaparral spread rate from
CAWFE (\ref{eq:chapparal}), and the explicit reduction of wind from 6.1\,m
height to midflame height, following \citet{Baughman-1980-EMW}.

In either case, the spread rate can be written as
\begin{equation} \label{eq:spread-WRF-Fire}
S=\max\left\{ S_0, R_0 + c\min\left\{ e,\max\left\{ 0,U\right\} \right\}^b +
d\max\left\{ 0,\tan\phi\right\}^{2} \right\} ,
\end{equation}
where $S_0$, $R_0$, $b$, $c$, $d$, $e$ are the fuel-dependent coefficients
that represent the spread rate internally. These coefficients are stored for
every grid point.

At a point on the fireline, denote by $\mathbf{n}$ the outside normal to the
fire region, $\mathbf{U}$ the wind vector, and $z$ the terrain height. The
normal component of the wind vector, $U=\mathbf{U}\cdot\mathbf{n}$, and the
normal component of the terrain gradient, $\tan\phi=\nabla z\cdot\mathbf{n}$,
are used to determine the spread rate, which is interpreted as the spread rate
in the normal direction $\mathbf{n}$.

\subsection{Fuel burned and heat released}

Each location starts with fuel fraction $F=1$. Once the fuel is ignited at a
time $t_{\rm i}$, the fuel fraction decreases exponentially,
\begin{equation} \label{eq:fuel}
F\left( t\right) =\exp\left( -\frac{\left( t-t_{\rm i} \right)}{T_{\rm f}}
\right), \quad t > t_{\rm i} ,
\end{equation}
where $t$ is the time, $t_{\rm i}$ is the ignition time, $F_0$ is the initial
amount of fuel, and $T_{\rm f}$ is the fuel burn time, i.e., the number of
seconds for the fuel to burn down to $1/e\approx 0.3689$ of the original
quantity. Since by definition of the fuel weight $w$ (Table~\ref{tab:fuels}),
the fuel burns down to 0.6 of the original quantity in 600\,s when $w=1000$,
we have
\[
0.6^{\frac{\left( t-t_{\rm i} \right)}{600} \frac{1000}{w}} = \exp\left( -
\frac{ \left( t-t_{\rm i} \right)}{T_{\rm f}}\right) ,
\]
which gives
\[
T_{\rm f} = -\frac{600 w}{1000\ln 0.6}\approx\frac{w}{0.8514}.
\]
The input coefficient $w$ is used in \mbox{WRF-Fire} rather than $T_{\rm f}$
for compatibility with existing fuel models and literature.

The average sensible heat flux density released in time interval $\left( t, t
+ \Delta t\right)$ is computed as
\begin{equation} \label{eq:sensible-heat-flux}
\phi_h = \frac{F\left( t \right) - F\left( t+\Delta t \right)}{\Delta t}
\frac{1}{1+M_{\rm f}}w_{\ell}h,\quad \left({\rm J\;m}^{-2}\;{\rm
s}^{-1}\right)
\end{equation}
and the average latent heat (i.e., moisture) flux density is given by
\begin{equation} \label{eq:latent-heat-flux}
\phi_{q}=\frac{F\left( t\right) -F\left( t+\Delta t\right) }{\Delta
t}\frac{M_{f}+0.56}{1+M_{f}}Lw_{\ell},\quad \left({\rm J\;m}^{-2}\;{\rm
s}^{-1}\right)
\end{equation}
where 0.56 is the estimated mass ratio of the water output from the combustion to
the dry fuel, and $L=2.5\times 10^6$\,J\,kg$^{-1}$ is the specific latent
heat of condensation of water at 0\,$^\circ$C, used for nominal conversion of
moisture to heat. This computation is from CAWFE.

\section{Domain, grids, and nodes}

\begin{figure}
\vspace*{2mm} \center\includegraphics[width=10cm]{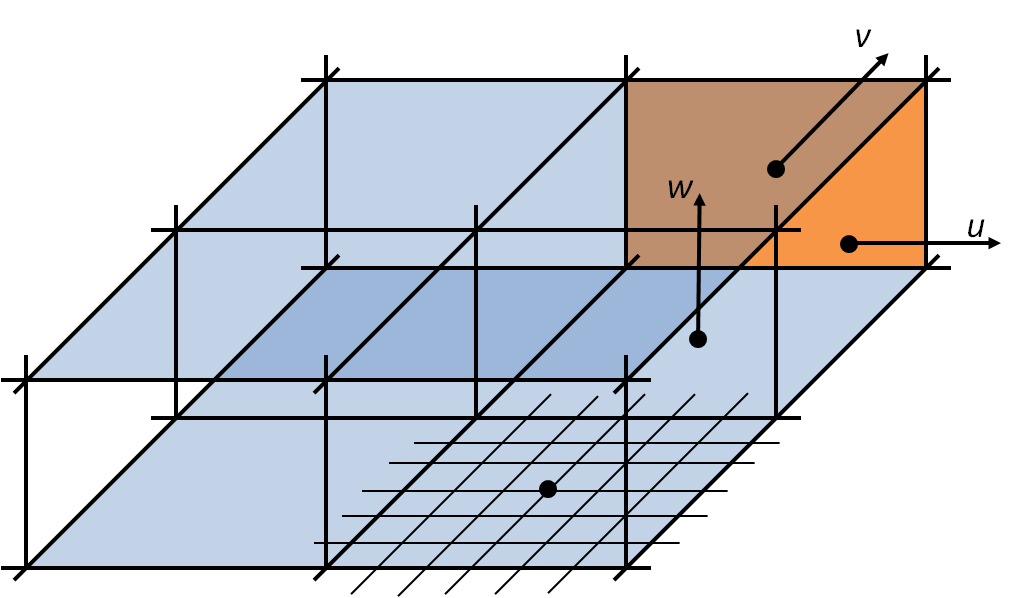}
\caption{One $2\times 2$ tile with the lowest layer of the atmospheric grid
and the fire subgrid on the surface shown. Wind vector components $u$, $v$,
$w$ are located at the midpoints of the sides of the atmospheric grid cells.
Some faces are colored for perspective. \label{fig:subgrid}}
\end{figure}

The atmospheric model operates on a logically quadrilateral \mbox{3-D} grid
on the Earth surface, and uses a sequence of horizontally nested grids,
called domains \citep{Kalnay-AMD-2003}. Only the innermost (the finest)
atmospheric domain is coupled with the fire model; see also
Sect.~\ref{sec:settings}. Scalar variables in the atmospheric model are
located at the centers of the 3D grid cells, while the wind vector components
are at a staggered grid at the midpoints of the cell faces. The fire model
operates on a refined fire mesh called the subgrid (Fig.~\ref{fig:subgrid}),
and all of its variables are all represented by their values at the centers
of the cells of this fire subgrid.

\section{Mathematical core of the fire model \label{sec:core}}

Subsections \ref{sec:levelset} and \ref{sec:fuel-fraction} below follow
\cite{Mandel-2009-DAW}.

\subsection{Fire propagation by the level-set method \label{sec:levelset}}

The model maintains a level-set function $\psi$, the time of ignition $t_{\rm
i}$, and the fuel fraction $F$. Denote a point on the surface by $\mathbf{x}
= \left( x, y \right)$. The burning region at time $t$ is represented by a
level-set function $\psi = \psi\left( \mathbf{x}, t \right) $ as the set of
all points $\mathbf{x}$ such that $\psi\left( \mathbf{x}, t \right) \leq 0$.
There is no fire at $\mathbf{x}$ if $\psi\left( \mathbf{x}, t \right) >0$.
The fireline is the set of all points $\mathbf{x}$ such that $\psi\left(
\mathbf{x}, t \right) =0$. Since on the fireline, the tangential component of
the gradient $\nabla\psi$ is zero, the outside normal vector at the fireline
is
\begin{equation} \label{eq:normal}
\mathbf{n}=\frac{\nabla\psi}{\left\Vert \nabla\psi\right\Vert }.
\end{equation}

Now consider a point $\mathbf{x}\left( t\right)$ that moves with the
fireline. Then the fire spread rate $S$ at $\mathbf{x}$ in the direction of
the normal $\mathbf{n}$ is
\begin{equation} \label{eq:normal-spread}
S=\mathbf{n}\cdot\frac{\partial\mathbf{x}}{\partial t},
\end{equation}
and, from the definition of the fireline, $\psi\left( \mathbf{x} \left( t
\right), t \right) =0$. By the chain rule and substituting from
(\ref{eq:normal}) and (\ref{eq:normal-spread}), we have
\begin{equation} \label{eq:chain-rule-fireline}
0=\frac{d}{dt}\psi\left( \mathbf{x},t\right) =\frac{\partial\psi}{\partial
t}+\frac{\partial\psi}{\partial x}\frac{\partial x}{\partial t}+\frac
{\partial\psi}{\partial y}\frac{\partial y}{\partial t}=\frac{\partial\psi
}{\partial t}+\left\Vert \nabla\psi\right\Vert \left( \mathbf{n}\cdot
\frac{\partial\mathbf{x}}{\partial t}\right) =\frac{\partial\psi}{\partial
t}+S\left\Vert \nabla\psi\right\Vert .
\end{equation}
So, the evolution of the level-set function is governed by the partial
differential equation
\begin{equation} \label{eq:levelset}
\frac{\partial\psi}{\partial t}+S \left\Vert \nabla\psi\right\Vert =0,
\end{equation}
called the \emph{level-set equation} \citep{Osher-2003-LSM}. The spread rate
$S$ is evaluated from (\ref{eq:spread-WRF-Fire}) for all $\mathbf{x}$, not
just on the fireline. Since $S\geq 0$, the level-set function does not
increase with time, and the fire area cannot decrease, which also helps with
numerical stability by eliminating oscillations of the level-set function
$\psi$ in time.

The level-set equation is discretized on a rectangular grid with spacing
$\left( \triangle x, \triangle y\right)$, called the fire grid. The level-set
function $\psi$ and the ignition time $t_{\rm i}$ are represented by their
values at the centers of the fire grid cells. This is consistent with the
fuel data given in the center of each cell also.

To advance the fire region in time, we use Heun's method (Runge-Kutta method
of order 2),
\begin{eqnarray} \label{eq:Heun}
\psi^{n+1/2} &  = & \psi^n + \Delta t F\left(  \psi^n \right) \nonumber\\[2mm]
\psi^{n+1} & = & \psi^n + \Delta t\left( \frac{1}{2} F \left( \psi^n \right)
+ \frac{1}{2} F\left( \psi^{n+1/2}\right) \right) ,
\end{eqnarray}
The right-hand side $F$ is a discretization of the term $-S\left\Vert
\nabla\psi\right\Vert$ with upwinding and artificial viscosity,
\begin{equation} \label{eq:F-disc}
F\left( \psi\right) = -S\left( \mathbf{U} \cdot \mathbf{n}, \nabla z \cdot
\mathbf{n} \right) \left \Vert \overline{\nabla} \psi \right\Vert +
\varepsilon \widetilde{\triangle}\psi,
\end{equation}
where $\mathbf{n}=\nabla\psi/\Vert\nabla\psi\Vert$ is computed by finite
central differences and $\overline{\nabla}\psi= \left[\, \overline{\nabla}_x
\psi, \overline{\nabla}_y \psi\, \right]$ is the upwinded finite difference
approximation of $\nabla\psi$ by Godunov's method
\citep[p.~58]{Osher-2003-LSM},
\begin{eqnarray} \label{eq:Godunov}
\overline{\nabla}_x \psi=\left\{
\begin{array}{lll}
                 \overline{\nabla}_x^+ \psi & ~~ {\rm if} & ~~ \overline{\nabla}_x^- \psi \leq 0 ~~ {\rm and} ~~ \overline{\nabla}_x^+ \psi\leq 0,                      \\[2mm]
                 \overline{\nabla}_x^- \psi & ~~ {\rm if} & ~~ \overline{\nabla}_x^- \psi \geq 0 ~~ {\rm and} ~~ \overline{\nabla}_x^+ \psi\geq 0,                      \\[2mm]
0                                           & ~~ {\rm if} & ~~ \overline{\nabla}_x^- \psi \leq 0 ~~ {\rm and} ~~ \overline{\nabla}_x^+ \psi\geq 0,                      \\[2mm]
{\rm otherwise}~~  \overline{\nabla}_x^- \psi & ~~ {\rm if} & ~~ \left\vert\overline{\nabla}_x^- \psi\right\vert \geq \left\vert \overline{\nabla}_x^+\psi\right\vert ,   \\[2mm]
\overline{\nabla}_x^+ \psi & ~~ {\rm if} & ~~ \left\vert
\overline{\nabla}_x^- \psi \right\vert \leq\left\vert \overline{\nabla}_x^+
\psi\right\vert ,
\end{array}
\right.
\end{eqnarray}
where $\nabla_x^+\psi$ and $\nabla_x^-\psi$ are the right and left one-sided
finite differences
\begin{eqnarray*}
\nabla_x^+ \psi\left( x, y \right) & = & \frac{\psi\left( x+ \triangle x, y\right)  - \psi\left( x, y \right)}{\triangle x},\\[2mm]
\nabla_x^- \psi\left( x, y \right) & = & \frac{\psi\left( x, y\right) - \psi
\left( x-\triangle x,y\right) }{\triangle x},
\end{eqnarray*}
and similarly for $\nabla_y^+\psi$ and $\nabla_y^-\psi$. Further, in
(\ref{eq:F-disc}), $\varepsilon$ is scale-free artificial viscosity
($\varepsilon=0.4$ here), and
\begin{eqnarray*}
\widetilde{\triangle}\psi & = & \nabla_x^+ \psi-\nabla_x^- \psi + \nabla_y^+ \psi-\nabla_y^- \psi     \\[2mm]
& = & \frac{\psi\left( x + \triangle x,y\right) - 2\psi\left( x, y\right) +
\psi\left( x-\triangle x, y \right)}{\triangle x} + ~~ {\rm similar~term~for}
~~ y
\end{eqnarray*}
is the five-point Laplacian of $\psi$ scaled so that the artificial viscosity
is proportional to the mesh step,
\[
\widetilde{\triangle}\psi\approx\triangle x\frac{\partial^{2}\psi}{\partial
x^2} + \triangle y \frac{\partial^2 \psi}{\partial y^2}.
\]

A numerically stable scheme with upwinding, such as (\ref{eq:Godunov}), is
required to compute the term $\left\Vert \nabla\psi\right\Vert$ in the level
set equation (\ref{eq:levelset}). However, in our tests, the gradient by
standard central differences,
\[
\nabla\psi\approx\left[  \frac{\psi\left(  x+\triangle x,y\right)
-\psi\left(  x-\triangle x,y\right)  }{2\triangle x},\frac{\psi\left(
x,y+\triangle y\right)  -\psi\left(  x,y-\triangle y\right)  }{2\triangle
y}\right]  ,
\]
worked better in the computation of the normal vector $\mathbf{n}$ by
(\ref{eq:normal}), which is used to evaluate the normal component of the wind
and the slope in (\ref{eq:spread-WRF-Fire}).

Before computing the finite differences up to the boundary, the level-set
function is extrapolated to one layer of nodes beyond the boundary. However,
the extrapolation is not allowed to decrease the value of the level-set
function to less than the value at either of the points it is extrapolated
from. For example, when $\left( i, j \right)$ is the last node in the domain
in the direction $x$, the extrapolation
\[
\psi_{i+1,j}=\max\left\{ \psi_{ij}+\left( \psi_{ij}-\psi_{i-1,j}\right),
\psi_{ij}, \psi_{i-1,j}\right\} ,
\]
is used, and similarly in the other cases. This is needed to avoid numerical
instabilities at the boundary. Otherwise, a decrease in $\psi$ at a boundary
node, which may happen with non-homogeneous fuels in real data, is amplified
by the extrapolation, and $\psi$ keeps decreasing at that boundary node in
every time step until it becomes negative, starting a spurious fire.

The model does not support fire crossing the boundary of the domain. When
$\psi<0$ is detected near the boundary, the simulation terminates. This is not
a limitation in practice, because the fire should be well inside the domain
anyway for a proper response of the atmosphere.

\subsection{Computation of the ignition time}

The ignition time $t_{\rm i}$ in the strip that the fire has moved over in
one time step is computed by linear interpolation from the level-set
function. Suppose that the point $\mathbf{x}$ is not burning at time $t$ but
is burning at time $t+\triangle t$, that is, $\psi\left( \mathbf{x},t\right)
>0$ and $\psi\left(  \mathbf{x},t+\triangle t\right)  \leq 0$. The ignition
time at $\mathbf{x}$ satisfies $\psi\left( \mathbf{x}, t_{\rm i} \left(
\mathbf{x} \right) \right) =0$. Approximating $\psi$ by a linear function in
time, we have
\[
\frac{\psi\left( \mathbf{x}, t_{\mathrm{i}}\right) -\psi\left(
\mathbf{x},t\right) }{t_{\mathrm{i}}\left( \mathbf{x}\right) -t}
\approx\frac{\psi\left( \mathbf{x},t+\triangle t\right) -\psi\left(
\mathbf{x},t_{\mathrm{i}}\right) }{t+\triangle t-t_{\mathrm{i}}\left(
\mathbf{x}\right) },
\]
and we take
\begin{equation} \label{eq:set-ignition-time}
t_{\rm i} \left(\mathbf{x}\right)=t+\frac{\psi\left( \mathbf{x},t\right)
\triangle t}{\psi\left( \mathbf{x}, t\right) -\psi\left(
\mathbf{x},t+\triangle t\right)}.
\end{equation}

\subsection{Computation of the fuel fraction \label{sec:fuel-fraction}}

The fuel fraction is approximated over each fire mesh cell $C$ by integrating
(\ref{eq:fuel}) over the fire region. Hence, the fuel fraction remaining in
cell $C$ at time $t$ is given by
\begin{equation} \label{eq:fuel_left}
F=1-\frac{1}{\operatorname*{area}\left( C\right) }{\iint\limits_{
\genfrac{}{}{0pt}{}{\mathbf{x}\in C}{\psi\left( \mathbf{x}, t\right) \leq 0}
}}1-\exp\left( -\frac{t-t_{\rm i} \left( \mathbf{x}\right) } {T_{\rm
f}(\mathbf{x})}\right) d\mathbf{x} .
\end{equation}
Once the fuel fraction is known, the heat fluxes are computed from
(\ref{eq:sensible-heat-flux}) and (\ref{eq:latent-heat-flux}). This scheme
has the advantage that the total heat released in the atmosphere over time is
exact, regardless of approximations in the computation of the integral
(\ref{eq:fuel_left}). Our objective in the numerical evaluation of
(\ref{eq:fuel_left}) is a method that is second order accurate when the whole
cell is on fire, exact when no part of the cell $C$ is on fire (namely,
returning the value one), and provides a natural transition between these two
cases. Just like standard schemes in numerical analysis can be derived from
the requirement that they are exact for all polynomials up to a given degree,
the guiding principle here is that the scheme should be exact in as many
special cases as possible. Then we expect that the scheme should work well
overall.

\begin{figure}
\vspace*{2mm} \center\includegraphics[width=8cm]{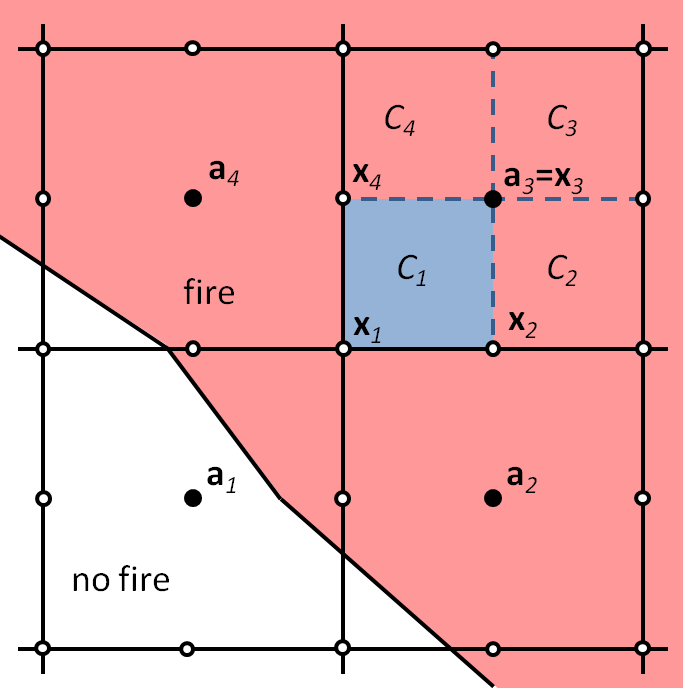}
\caption{Division of fire mesh cells into subcells for fuel fraction
computation. The level-set function $\psi$ and the ignition time $t_{\rm i}$
are given at the centers $\mathbf{a}_1,\ldots,\mathbf{a}_4$ of the cells of
the fire grid. The integral (\ref{eq:fuel_left_subcells}) over the cell $C$
with the center $\mathbf{a} _{3}$ is computed as the sum of integrals over
the subcells $C_1,\ldots,C_4$. While the values of $\psi$ and $t_{\rm i}$ are
known at $\mathbf{a}_3=\mathbf{x}_3$, they need to be interpolated to the
remaining corners $\mathbf{x}_1$, $\mathbf{x}_2$, $\mathbf{x}_4$ of the
subcell $C_1$ from their values at the points
$\mathbf{a}_1,\ldots,\mathbf{a}_4$. \label{fig:subcells}}
\end{figure}

While the fuel burn time $T_{\rm f}$ can be interpolated as constant over the
whole cell, the level-set function $\psi$ and the ignition time $t_{\rm i}$
must be interpolated more accurately to allow a submesh representation of the
burning area and a gradual release of the heat as the fireline moves over the
cell. In addition, we need the fuel fraction computed over each mesh cell,
because the heat fluxes in the mesh cells are summed up to give the heat flux
in an atmospheric cell. Our solution is to split each cell into 4\,subcells
$C_j$, interpolate to the corners of the subcells, and add the integrals,
\begin{equation} \label{eq:fuel_left_subcells}
{\iint\limits_{%
\genfrac{}{}{0pt}{}{\mathbf{x}\in C}{\psi\left( \mathbf{x}\right) \leq0}
}1-}\exp\left( -\frac{t-t_{\rm i} \left( \mathbf{x}\right) } {T_{\rm
f}(\mathbf{x})}\right) d\mathbf{x=}\sum_{j=1}^{4}{\iint\limits_{
\genfrac{}{}{0pt}{}{\mathbf{x}\in C_j}{\psi\left( \mathbf{x}\right) \leq0}
}1-}\exp\left( -\frac{t-t_{\rm i} \left( \mathbf{x}\right) } {T_{\rm
f}\left(\mathbf{x}\right)}\right) d\mathbf{x},
\end{equation}
cf., Fig.~\ref{fig:subcells}. The level-set function $\psi$ is interpolated
bilinearly to the vertices of the subcells $C_j$, and the burn time $T_{\rm
f}$ is constant on each $C_j$, given by its value at the fire grid nodes.
However, to interpolate the ignition time $t_{\rm i}$ we first define $t_{\rm
i}$ outside of the fire region and on the fireline by
\begin{equation} \label{eq:ti-comp}
t_{\rm i} = t ~~ {\rm if} ~~ \psi\geq 0.
\end{equation}
This allows us to omit the condition $\psi\leq 0$ in the definition of the
integration domains in (\ref{eq:fuel_left_subcells}) and integrate on the
whole cells, respective subcells, only. Then, we interpolate $t_{\rm i}$
bilinearly to the vertices of the subcells $C_j$ and correct the resulting
values by applying the compatibility condition (\ref{eq:ti-comp}).

To compute the integral over a subcell $C_j$, we first estimate the fraction
of the subcell that is burning, by
\begin{equation} \label{eq:approx-area}
\frac{\operatorname*{area}\left\{ \mathbf{x}\in C_j:\psi \left( \mathbf{x}
\right) \leq 0 \right\} }{\operatorname*{area}(C_j)}\approx \beta =
\frac{1}{2} \left( 1-\frac{\sum_{k=1}^4 \psi\left( \mathbf{x}_k \right)
}{\sum_{k=1}^4 \left\vert \psi\left( \mathbf{x}_k \right) \right\vert}\right)
,
\end{equation}
where $\mathbf{x}_k$ are the the corners of the subcell $C_j$. This
approximation is exact when no part of the subcell $C_j$, is on fire, that
is, all $\psi\left( \mathbf{x}_k\right) \geq0$ and at least one $\psi\left(
\mathbf{x}_k \right) >0$; the whole $C_j$ is on fire, that is, all
$\psi\left( \mathbf{x}_k \right) \leq0$ and at least one $\psi\left(
\mathbf{x}_k \right) < 0$; or the values $\psi\left( \mathbf{x}_k \right)$
define a linear function and the fireline crosses the subcell diagonally or
it is aligned with one of the coordinate directions.

Next, replace $t_{\rm i} \left( \mathbf{x}_k \right)$ by $t$ when $\psi\left(
\mathbf{x}_k \right) >0$ (i.e., the node $\mathbf{x}_k$ is not on fire), and
compute the approximate fraction of the fuel burned as
\begin{equation} \label{eq:approx-fraction}
\frac{1}{\operatorname*{area}\left(  C\right)  }{\iint\limits_{%
\genfrac{}{}{0pt}{}{\mathbf{x}\in C}{\psi\left(  \mathbf{x},t\right)  \leq0}%
}}1-\exp\left( -\frac{t-t_{\rm i} \left( \mathbf{x}\right) } {T_{\rm f}
\left(\mathbf{x} \right)}\right) d \mathbf{x\approx}\beta\left( 1-\exp\left(
-\frac{1}{4} \sum_{k=1}^4\frac{t_{\rm i} \left( \mathbf{x}_k\right)
-t}{T_{\rm f}}\right) \right)
\end{equation}
This calculation is accurate asymptotically when the fuel burns slowly and
the approximation $\beta$ of the burning area is exact.

\subsection{Ignition \label{sec:ignition}}

Typically, a fire starts from a horizontal extent much smaller than the fire
mesh cell size, and both point and line ignition need to be supported. The
previous ignition mechanism \citep{Mandel-2009-DAW} ignited everything within
a given distance from the ignition line at once. This distance was required
to be at least one or two mesh steps, so that the initial fire is visible on
the fire mesh, and the fire propagation algorithm from
Sect.~\ref{sec:levelset} can catch on. This caused an unrealistically large
initial heat flux and the fire started too fast.

The current ignition scheme achieves submesh resolution and zero-size
ignition. A small initial fire is superimposed on the regular propagation
mechanism, which then takes over. Drip-torch ignition is implemented as a
collection of short ignition segments that grows at one end every time step.
Multiple ignition segments are also supported.

The model is initialized with no fire by choosing the level-set function
$\psi\left( \mathbf{x}, t_0 \right){=}\operatorname*{const}>0$. Consider an
initial fire that starts at time $t_{\rm g}$ on a segment
$\overline{\mathbf{a, b}}$ and propagates in all directions with an initial
spread rate $S_{\rm g}$ until the distance $r_{\rm g}$ is reached. At the
beginning of every time step $t$ such that
\[
t_{\rm g} \leq t\leq t_{\rm g} + \frac{r_{\rm g}}{S_{\rm g}},
\]
we construct the level-set function of the initial fire,
\begin{equation} \label{eq:ignition-lfn}
\psi_{\rm g} \left( \mathbf{x}, t \right) =\operatorname*{dist}\left(
\mathbf{x}, \overline{\mathbf{a,b}}\right) -S_{\rm g} \left( t - t_{\rm g}
\right)
\end{equation}
and replace the level-set function of the model by
\begin{equation} \label{eq:ignition-min}
\psi\left( \mathbf{x}, t \right) := \min\left\{ \psi\left( \mathbf{x}, t
\right), \psi_{\rm g} \left( \mathbf{x}, t \right) \right\} .
\end{equation}
For a drip-torch ignition starting from point $\mathbf{a}$ at time $t_{\rm
g}$ at velocity $\mathbf{v}$ until time $t_{\rm h}$, the ignition line at
time $t$ is the segment $\overline{\mathbf{a},\mathbf{a} +\mathbf{v}\left(
\min\left\{ t, t_{\rm h} \right\} - t_{\rm g}\right)}$, and
(\ref{eq:ignition-lfn}) becomes
\[
\psi_{\rm g} \left( \mathbf{x}, t\right) =\operatorname*{dist}\left(
\mathbf{x}, \overline{\mathbf{a}, \mathbf{a} + \mathbf{v}\left( \min\left\{
t, t_{\rm h} \right\} -t_{\rm g} \right)}\right) -\min\left\{ r_{\rm g},
S_{\rm g} \left( t-t_{\rm g} \right) \right\}
\]
followed again by (\ref{eq:ignition-min}), at the beginning of every time
step begining at time $t$ such that
\[
t_{\rm g} \leq t\leq t_{\rm h} + \frac{r_{\rm g}}{S_{\rm g}}.
\]
The ignition time of newly ignited nodes is set to the arrival time of the
fire at the spread rate $S_{\rm g}$ from the nearest point on the ignition
segment.

\section{Atmospheric model \label{sec:WRF}}

We summarize some background information about WRF-ARW from
\citet{Skamarock-2008-DAR}, to the extent needed to understand the coupling
with the fire module.

The model is formulated in terms of the hydrostatic pressure vertical
coordinate $\eta$, scaled and shifted so that $\eta=1$ at the Earth surface
and $\eta=0$ at the top of the domain. The governing equations are a system of
partial differential equations of the form
\begin{equation} \label{eq:WRF}
\frac{d\Phi}{dt}=R\left( \Phi\right) ,
\end{equation}
where $R$ contains also the advection terms, and $\Phi=\left( U, V, W,\phi',\Theta,\mu', Q_{\rm m}\right) $. The
fundamental WRF variables are $\mu=\mu\left( x,y\right)$, the hydrostatic
component of the pressure differential of dry air between the surface and the
top of the domain, written in perturbation form $\mu =\overline{\mu}+\mu'$,
where $\overline{\mu}$ is a reference value in hydrostatic balance; $U=\mu
u$, where $u=u\left( x,y,\eta\right)$ is the Cartesian component of the wind
velocity in the $x$-direction, and similarly $V$ and $W$; $\Theta=\mu\theta$,
where $\theta=\theta\left( x,y,\eta\right)$ is the potential temperature;
$\phi=\phi\left( x,y,\eta\right) =\overline{\phi}+\phi'$ is the geopotential;
and $Q_{\rm m}=\mu q_{\rm m}$ is the moisture content of the air. The
variables in the state $\Phi$ evolved by (\ref{eq:WRF}) are called prognostic
variables. Other variables computed from them, such as the hydrostatic
pressure $p$, the thermodynamic temperature $T$, and the height $z$, are
called diagnostic variables. The variables that contain $\mu$ are called
coupled. The value of the right-hand side $R\left( \Phi\right)$ is called
tendency. See \citet[p.~7--13]{Skamarock-2008-DAR} for details and the form
of $R$.

The system (\ref{eq:WRF}) is discretized in time by the explicit 3rd order
Runge-Kutta method
\begin{eqnarray} \label{eq:rk3}
\Phi_1            & = & \Phi^t + \frac{\Delta t}{3}R\left(  \Phi^{t}\right) \nonumber\\
\Phi_2            & = & \Phi^t + \frac{\Delta t}{2}R\left(  \Phi_{1}\right) \nonumber\\
\Phi^{t+\Delta t} & = & \Phi^t + \Delta tR\left( \Phi_{2}\right)
\end{eqnarray}
where the differential operator $R$ is discretized by finite differences and
the tendencies from physics packages, such as the fire module, are updated
only the third Runge-Kutta step \cite[p.~16]{Skamarock-2008-DAR}. In order
to avoid small time steps, the tendency in the third Runge-Kutta step also
includes the effect of substeps to integrate acoustic modes.

\section{Coupling of the fire and the atmospheric models  \label{sec:coupling}}

The terrain gradient is computed from the terrain height at the best
available resolution and interpolated to the fire mesh in preprocessing.
Interpolating the height and then computing the gradient would cause jumps in
the gradient, which affect fire propagation, unless high-order interpolation
is used.

In each time step of the atmospheric model, the fire module is called from
the third step of the Runge-Kutta method. First the wind is interpolated to a
given height $z_{f}$ above the terrain (currently, 6.1\,m following BEHAVE),
assuming the logarithmic wind profile
\[
u\left( z \right) \approx\left\{
\begin{array}{lr}
\operatorname*{const}\, \ln\frac{z}{z_{0}}, &        z \geq z_{0},\\
0                                         & 0 \leq z\leq z_{0},
\end{array}
\right.
\]
where $z$ is the height above the terrain and $z_0$ is the roughness height.
For a fixed horizontal location, denote by $z_1$, $z_2,\ldots$ the heights of
the centers of the atmospheric mesh cells; these are computed from the
geopotential $\phi$, which is a part of the solution. The horizontal wind
component $u\left( z_{f}\right)$ under the $u$-points
(Fig.~\ref{fig:subgrid}) is then found by vertical log-linear interpolation,
that is, $u\left( z_{f}\right)$ is found by \mbox{1-D} piecewise linear
interpolation of the values $u\left( z_0 \right) =0$, $u\left( z_1 \right)$,
$u\left( z_2 \right),\ldots$ at $\ln z_0$, $\ln z_1$, $\ln z_2,\ldots$ to
$\ln z_{\rm f}$. If $z_{\rm f}\leq z_0$, we set $u\left( z_{\rm f}\right)
=0$. The $v$ component of the wind is interpolated vertically in the same
way. Each horizontal wind component $u$, $v$ is then interpolated separately
to the cell centers of the fire subgrid by bilinear interpolation.

The fire model then makes one time step:
\begin{enumerate}
\item If there are any active ignitions, the level-set function is updated and
the ignition times of any newly ignited nodes are set following
Sect.~\ref{sec:ignition}.
\item The numerical scheme (\ref{eq:Heun})--(\ref{eq:Godunov}) for the level
set equation (\ref{eq:levelset}) is advanced to the next time step.
\item The time of ignition set for any any nodes that were ignited during the
time step, from (\ref{eq:set-ignition-time}).
\item The fuel fraction is updated following Sect.~\ref{sec:fuel-fraction}.
\item The sensible and latent heat flux densities are computed from
(\ref{eq:sensible-heat-flux}) and (\ref{eq:latent-heat-flux}) in each fire
model cell.
\item The resulting heat flux densities are averaged over the fire cells that
make up one atmosphere model cell, and inserted into the atmospheric model,
which then completes its own time step.
\end{enumerate}

The heat fluxes from the fire are inserted into the atmospheric model as
forcing terms in the differential equations of the atmospheric model into a
layer above the surface, with assumed exponential decay with altitude. Such
scheme is needed because WRF does not support flux boundary conditions. This
is code originally due to \citet{Clark-1996-CAF,Clark-1996-CAM} and it was
rewritten for WRF variables in \citet{Patton-2004-WCA}. The sensible heat
flux is inserted as an additional source term to the equation for the potential temperature $\theta$, equal
to the vertical divergence of the heat flux,
\[
\frac{d\left(  \mu\theta \right)}{dt} \left( x, y, z \right) = R_{\Theta}\left(
\Phi \right) + \frac{\mu\left( x, y \right) \phi_{h}\left( x, y\right)
}{\sigma\varrho\left( x, y, z\right)} \frac{\partial}{\partial z} \exp\left(
- \frac{z}{z_{\rm ext}}\right)  ,
\]
where $R_{\Theta}\left( \Phi \right)$ is the component of the tendency in the
atmospheric model equations (\ref{eq:WRF}), $\sigma$ is the specific heat of
the air, $\varrho\left( x, y, z\right)$ is the density, and $z_{\rm ext}$ is
the heat extinction depth, given as parameter \texttt{fire\_ext\_grnd} in
\texttt{namelist.input}. The latent heat flux is inserted similarly into the
tendency of the vapor concentration $q_{\rm m}$ by
\[
\frac{d\left( \mu q_{\rm m}\right)}{dt} \left( x, y, z \right) = R_{Q_{\rm
m}} \left( \Phi \right) + \frac{\mu\left( x, y \right) \phi_{q} \left( x, y
\right)}{L\varrho \left( x, y, z \right)} \frac{\partial}{\partial z}\exp
\left( -\frac{z}{z_{\rm ext}}\right),
\]
where $L$ is the specific latent heat of the air. Cf. \citet[Eqs.~10, 12,
13,~18]{Clark-1996-CAF}.

\section{Software structure \label{sec:software}}

\subsection{Parallel structure}

\begin{figure}
\vspace*{2mm} \center\includegraphics[width=8cm]{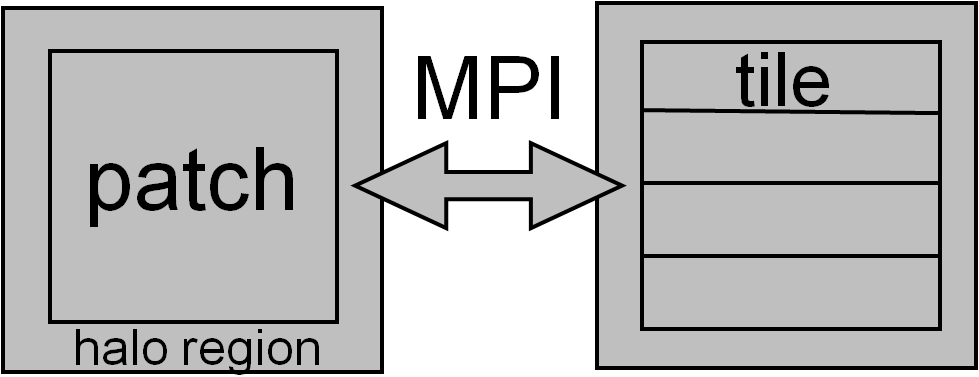}
\caption{Parallel communication in WRF. The computational domain is divided
into disjoint rectangular patches. Each patch is updated by a single MPI
process (distributed memory parallelism), and the process may read arary data
in a strip around the patch, called halo region. The communication between
the patches is by halo calls to the RSL parallel infrastructure
\citep{Michalakes-1999-RPR}, which update the halo regions by the values from
the neighboring patches. Each patch may be divided into tiles, which execute
in separate OpenMP threads (shared memory parallelism). Following WRF coding
conventions \citep{WRF-conventions}, computational kernels execute in a
single tile. They may read array values from a strip beyond the tile boundary
but no explicit communication is allowed. \mbox{3-D} arrays are divided into
patches and tiles in the horizontal plane, cf., Fig.~\ref{fig:subgrid}.
\label{fig:halo}}
\end{figure}

\begin{figure}
\vspace*{2mm} \center\includegraphics[width=12cm]{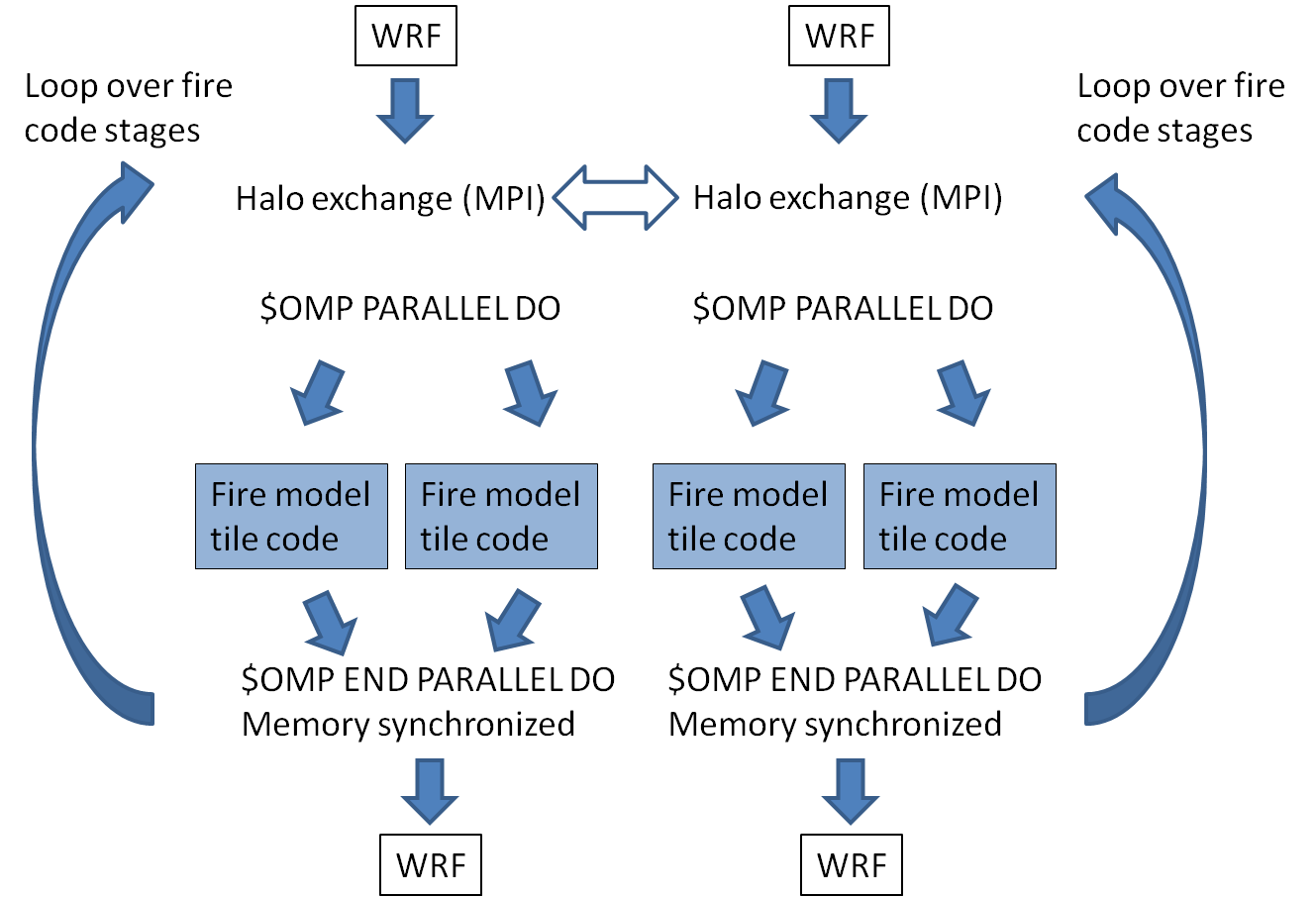}
\caption{Parallel structure of the fire module in the WRF physics layer. The
core code itself executes on a single tile, with all communication done
outside. Multiple passes through the fire module are needed in each time
step. \label{fig:parallel}}
\end{figure}

Parallel computing imposes a significant constraint on user programming
technique. WRF parallel infracture \citep{Michalakes-1999-RPR} divides the
domain horizontally into patches. Each patch executes in a separate MPI\
process and it may be further divided into tiles, which execute in separate
OpenMP threads (Fig.~\ref{fig:halo}). Communication between the tiles is
accomplished by exiting the OpenMP parallel loop over the tiles. The fire
grid tiles are colocated with the atmospheric grid tiles
(Fig.~\ref{fig:subgrid}). The patches are declared in memory with larger
bounds than the patch size, and communication between the patches is
accomplished by HALO calls (actually, includes of generated code), which
update a layer of array entries beyond the patch boundary from other patches.
The fire module computational code itself is designed to be
\emph{tile-callable} as required by the WRF coding conventions
\citep{WRF-conventions}. Tile-callable code updates array values on a single
tile, assuming that it can safely read data from a layer of several array
entries beyond the tile boundary. The communication (OpenMP loops or HALO
calls) happens outside; this means that every time when communication is
needed, tile-callable code must exit, and then the next stage can resume on
the next call (Fig.~\ref{fig:parallel}). The fire module code executes in
6\,stages interleaved with communication, 3\,stages for initialization and
3\,stages in every time step.

\subsection{Software layers}

\begin{figure}
\vspace*{2mm} \center\includegraphics[width=12cm]{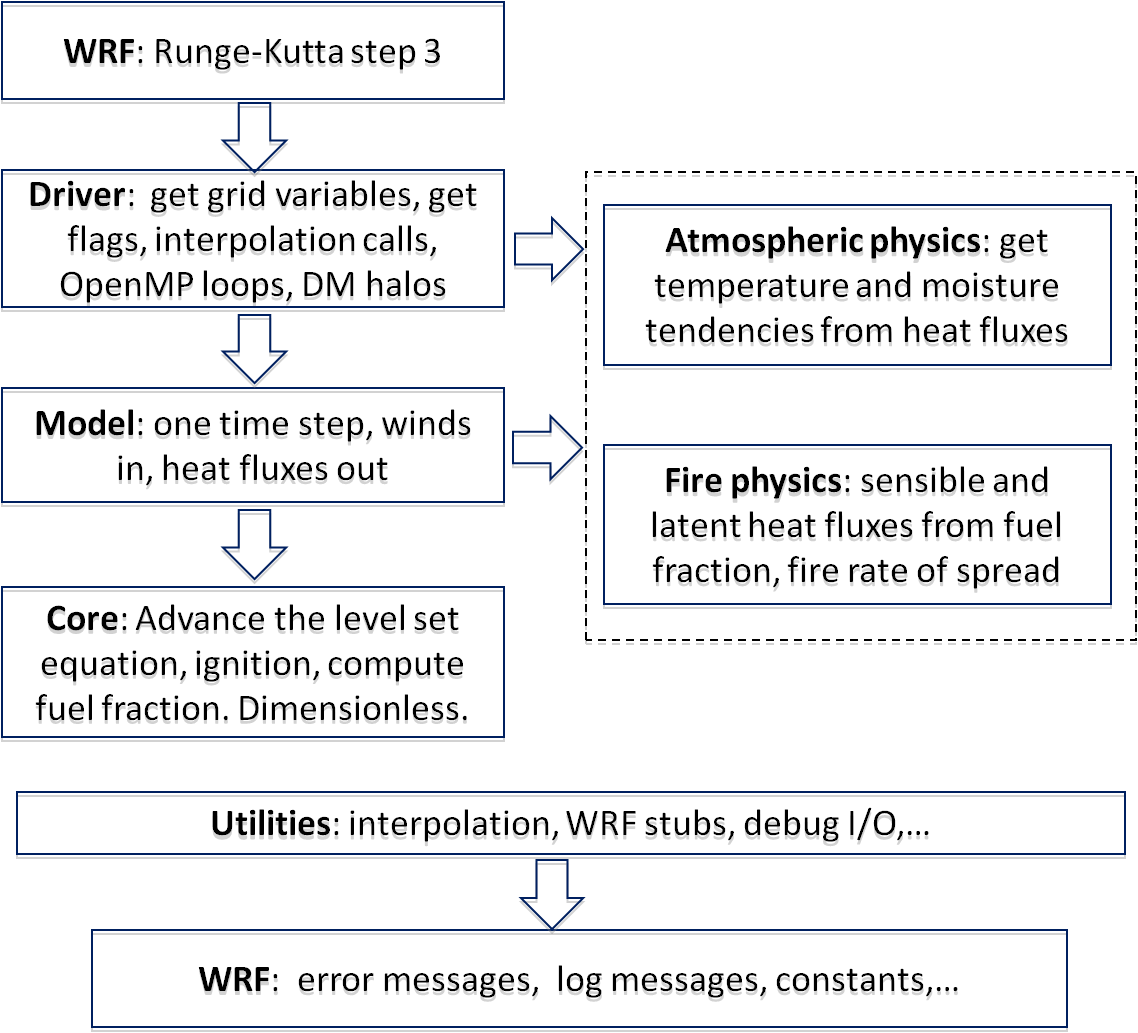}
\caption{Software layers of \mbox{WRF-Fire}. All physics dependencies are in
the dashed box. The utilities layer is called from all the other layers
above. \label{fig:structure}}
\end{figure}

The fire module software is organized in several isolated layers
(Fig.~\ref{fig:structure}). The \emph{driver layer} contains all exchange of
data between the tiles in parallel execution. The rest of the code is
tile-callable. The driver layer calls the interpolation and other coupling
between the fire and the atmospheric grids, and the fire code itself. The
\emph{atmospheric physics} layer mediates the insertion of the fire fluxes
into the atmosphere, as described in Sect.~\ref{sec:physical-model}. Only
these two layers depend on WRF; the rest of the fire module can be used as a
standalone code, independent on WRF. The \emph{utility layer} contains
interpolation and other service code, such as stubs to control access to WRF
infrastructure, so that WRF calls can be easily emulated in the standalone
code. The \emph{model layer} is the entry point to the fire module. The
\emph{core layer} is the engine of the fire model, described in
Sect.~\ref{sec:core}. The \emph{fire physics layer} evaluates the fire spread
rate and heat fluxes from fuel properties. One of the goals of the design is
that the only components that will need to be modified when the fire module
is connected to another atmospheric model in future are the driver layer, the
atmospheric physics layer, and the WRF stubs in the utility layer.

\section{Recommended WRF settings \label{sec:settings}}

\subsection{Domains and nesting}

WRF-Fire may be run in both ``ideal'' and ``real'' modes, which require
slightly different setups. In both cases, the model requires a set of data
defining model initialization (\texttt{wrfinput}). In the real cases,
boundary conditions in a form of \texttt{wrfbdy} files must be also provided,
and both types of files are created by \texttt{real.exe} preprocessor from
the WRF Preprocessing System (WPS). These files contain not only
meteorological and topographical data but also fire related information, such
as the fuel type map and high-resolution topography on the fire mesh. Since
the \mbox{WRF-Fire} initialization for the real cases does not differ from
the one for the regular WRF, all physical and dynamical options available in
the regular WRF are also available in \mbox{WRF-Fire}. Therefore, the same
general rules apply to the configuration of \mbox{WRF-Fire} as to the
configuration of the regular WRF. However, one should keep in mind that
resolutions of the finest domains in fire simulations are usually
significantly higher than in weather forecasting applications. This has two
consequences in terms of the proper \mbox{WRF-Fire} setup. First, if the
resolution of any of the inner domains is less than 100\,m, this domain
should be actually resolved in the large eddy simulation (LES) mode, without
the boundary layer parameterizations. At this resolution, the model should be
able to resolve the most energetic eddies responsible for mixing within the
boundary layer, so the boundary layer parameterization in this case is not
needed. Second, since in the nested mode, vertical levels are common for all
domains, the height of the first model level selected for the most outer
(parent) domain, defines also the level of the first model layer for all
inner (child) domains, even if their horizontal resolutions are an order of
magnitude smaller. The fact that the vertical model resolution is the same
for all domains significantly limits the minimum height above the ground of
the first model level. This in turn is crucial for the fire model, which uses
the wind speed interpolated to 6.1\,m above the ground. Therefore, in the
cases when the first model level must be relatively high above the ground it
is recommended to perform downscaling using the \texttt{ndown.exe} program,
being a part of the WRF distribution. In this case the outer domains are run
separately without the fire, and then based on the output from this
simulation, \texttt{ndown.exe} creates a set of new initial and boundary
condition files (\texttt{wrfinput} and \texttt{wrfbdy}) for the separate
simulation from the innermost domain(s). This allows for a new setup of
vertical levels for the innermost domains, and selecting proper physical
options for them.

\subsection{Large Eddy Simulation and surface properties}

To enable the high-resolution simulation in Large Eddy Simulation (LES) mode,
user should first disable the boundary layer parameterization
(\texttt{bl\_pbl\_physics=0}). The LES mode requires the proper surface fluxes
in order work properly. We recommend the option \texttt{isfflx=1}, which makes
WRF use a surface model to compute the surface fluxes. Other options with
constant heat fluxes and drag are not well suited for fire simulations. Out of
all surface exchange parameterizations only the classic Monin-Obukhov theory
(\texttt{sf\_sfclay\_physics=1}) is recommended for the LES cases. This option
assures a proper computation of surface transfer coefficients that are used
together with the surface properties (provided by the surface model) for
computation of the surface fluxes of the momentum, heat and moisture. The
surface model itself computes properties of the surface, but does not compute
the surface exchange coefficients, which are needed for computation of the
surface fluxes. Hence, in order to compute them, the surface properties must
be provided by a surface model, which is enabled by choosing a non-zero
\texttt{sf\_surface\_physics}. The subgrid scale parameterization used by the
WRF in LES mode is defined by the \texttt{km\_opt} parameter, which should be
set to 2 (TKE closure), or 1 (Smagorinsky scheme).

In real cases, \texttt{real.exe} automatically provides proper initialization
for the selected land surface model, and all other components. In idealized
cases, users have an option of the basic surface initialization, intended to
be used without the surface model, or the full surface initialization
(\texttt{sfc\_full\_init=1}). One should keep in mind that without the full
surface initialization, there is no direct way to define surface properties
such as temperature or roughness. For idealized cases with the full surface
initialization, the surface scheme utilizes a table containing records of
land-use categories and corresponding surface properties like roughness
length, heat capacity, etc. All these properties are defined in a text file
\texttt{LANDUSE.TBL}, which may be edited by the user. Therefore, setting up
the land-use category is enough to provide all static surface properties. The
basic parameters required by this surface model like land use index, surface
air temperature and soil temperature, may be defined directly in
\texttt{namelist.input} by the variables \texttt{sfc\_lu\_index},
\texttt{sfc\_tsk}, and \texttt{sfc\_tmn} if they are intended to be the same
over the whole domain. If they are not spatially uniform, they may be read in
from external files if \texttt{fire\_read\_lu}, \texttt{fire\_read\_tsk}, or
\texttt{fire\_read\_tmn} are set to true. For details about the input data
for real cases, see Sect.~\ref{sec:data-input}.

\subsection{Fire subgrid refinement ratios}

The fire mesh needs to be about 10\,times finer than the atmospheric mesh to
allow for gradual heat release into the atmosphere, even if fuel and
topography data may not be available at such fine resolution. The fire mesh
refinement in the $x$ and $y$ direction (\texttt{sr\_x} and \texttt{sr\_y})
must be defined in the domain section of \texttt{namelist.input}. Since these
refinement factors define dimensions of the fire-related variables, they must
be selected before execution of \texttt{real.exe}, which generates the WRF
input files. Any change in atmospheric to fire grid ratios requires
re-running \texttt{real.exe} and creating new input files. The atmospheric
mesh step should be about 60\,m or less for proper feedback of the wind on
the fire line; larger mesh step can result in incorrect fire spread rates and
atmospheric behavior \citep[p.~887]{Clark-1996-CAF}.

\subsection{Time step \label{sec:timestep}}

In real \mbox{WRF-Fire} simulations performed in multi-domain configurations
the time step requirements for the outer domains (run without fire) do not
differ from general meteorological cases. The recommended time step of
6\,times the horizontal grid spacing (in km) may be used as a starting point.
However, for the finest domains run with fire simulations, the time step in
most cases must be significantly smaller. For domains with low vertical
resolution and simple topography, the horizontal mesh step is crucial
for numerical stability, since the horizontal velocity is greater than the vertical one. In
fire simulations with high vertical resolution, the vertical velocity induced
by fire may violate the CFL condition. Therefore, it is advisable to use a
vertically stretched grid, with finer resolution at the surface (where
updraft velocities are not that high) and lower resolution at higher levels
where stronger updrafts are expected. This allows for having the first model
level relatively close to the ground, yet with vertical spacing aloft big
enough to handle strong convective updrafts without violating the CFL
condition.

In real cases, the pressure levels may be defined directly in the
\texttt{namelist.input} file. In ideal \mbox{WRF-Fire} runs, there is now an
option \texttt{stretch\_hyp}, which turns on hyperbolic grid stretching. The
grid refinement may be adjusted using the \texttt{z\_grd\_scale} namelist
variable. One should keep in mind that running the \mbox{WRF-Fire}
simulations with high-resolution topography in most cases limits the maximum
numerically stable time step. Steep terrain often induces high vertical
velocities that may violate the CFL condition. Therefore, these cases usually
require significantly smaller time steps than similar simulations run with
low-resolution, smooth topography.\vspace*{-4mm}

\section{Data input \label{sec:data-input}}\vspace*{-.5mm}

A WRF ideal run is used for simulations on artificial data. An additional
executable, \texttt{ideal.exe}, is run first to create the WRF input. A
different \texttt{ideal.exe} is built for each ideal case, and the user is
expected to modify the source of such ideal case to run custom experiments.
The ideal run for fire supports optional input of gridded arrays for land
properties, such as terrain height, roughness height, and terrain height.
This allows to run simulations which go beyond what would normally be
considered an ideal run and simplifies custom data input; the simulation of
the FireFlux experiment (Sect.~\ref{sec:simulations}) was done in this way.

A WRF real run is used for prediction and analysis of natural events. For a
real run, a user must supply data for the initial and boundary conditions for
the WRF simulation. The WRF Preprocessing System (WPS)
\cite[Chapter~3]{Wang-2010-AUG} contains a number of utilities useful for
preparing standard atmospheric and surface datasets for input into
WRF.\vspace*{-.5mm}
\begin{enumerate}
\item Geogrid creates the surface mesh from a specified geographic projection
and interpolates static surface data onto the mesh. It supports several
interpolation methods as well as data smoothing and creation of gradient
fields. Geogrid reads data in a tiled binary format described by a text file
and writes to a NetCDF file for each nested mesh. All data required for
atmospheric simulations up to 30\,arcseconds resolution globally are provided
by NCAR.
\item Ungrib extracts atmospheric data from standard GRIB files and writes to
a simple binary format. Ungrib does not do any interpolation; it only
searches through a number of files for necessary variables within the time
window of the simulation. Data for ungrib must be obtained by the user.
Several free sources of atmospheric GRIB data are available online from
production weather simulation.
\item Metgrid reads the output from geogrid and ungrib and produces a series
of NetCDF files read by WRF's \texttt{real.exe} binary. The geogrid output is
copied directly into each of these files, while the ungrib output is
interpolated horizontally on to the computational mesh.
\end{enumerate}

The metgrid files produced by WPS are portable and relatively compact so they
can be transferred to a computer cluster for the simulation's execution. From
this point, the \texttt{real.exe} program in WRF handles the vertical
interpolation of atmospheric fields and all processing for the creation of
WRF's initial (\texttt{wrfinput}) and boundary (\texttt{wrfbdy}) files.

WPS has been extended with the ability to produce data defined on the refined
surface meshes used by \mbox{WRF-Fire} (Sect.~\ref{sec:settings}); however,
it is not possible to distribute high resolution, global fields as is done in
the standard dataset. Instead, the user must download any necessary high
resolution fields and convert them into geogrid's binary format for each
simulation. \mbox{WRF-Fire} is distributed with an additional utility,
\texttt{convert\_geotiff.x}, which can perform this conversion from any
GeoTIFF file. This utility is written entirely in C and depends only on the
GeoTIFF library.

For a \mbox{WRF-Fire} simulation, it is only strictly necessary to download
one additional dataset for input into geogrid. This dataset contains fuel
behavior categories and is stored in the variable \texttt{NFUEL\_CAT}. For
simulations within the United States, this data can be obtained in GeoTIFF
format from the USGS at \url{http://www.landfire.gov}. \mbox{WRF-Fire} uses
an additional variable for topography, \texttt{ZSF}, which is allowed to be
different from the topography used used by the atmospheric code defined by
\texttt{HGT}. This is useful because a high resolution WRF simulation
generally requires the topography to be highly smoothed in preprocessing for
numerical stability. The fire code can benefit from a rougher topography for
more accurate fire spread computations.

Once the static data is converted into the geogrid binary format, the
\texttt{GEOGRID.TBL} should be edited to inform geogrid of the location of
each supplementary dataset. \mbox{WRF-Fire} expects two variables to be
created on the refined subgrid (\texttt{NFUEL\_CAT} and \texttt{ZSF}), this
is indicated by the line \texttt{subgrid=yes}; all other variables will be
defined on the standard atmospheric grid.

For atmospheric data, it is best to use the highest resolution dataset
available to initialize a \mbox{WRF-Fire} simulation to capture as much of
the local conditions near the fire as possible. Generally, publicly available
atmospheric data is limited to around 10\,km resolution. As a consequence,
one should create several nested grids, each with a 3 to 1 refinement ratio,
and a long spin-up prior to ignition in order to recreate local conditions.
Preliminary results indicate that assimilation of data from weather stations
or satellite radiances may be required for an accurate simulation
(\citealp{Beezley-2010-SMC}).\vspace*{-3mm}

\section{Computational simulations \label{sec:simulations}}

\citet{Kim-2011-RDE} has verified that the level-set method in the fire
module advects the fire shape correctly, on some of the same examples that
were used to verify the tracer code in CAWFE \citep{Clark-2004-DCA}.

A number of successful simulations with \mbox{WRF-Fire} now exist.

\begin{figure}
\vspace*{2mm} \center\includegraphics[width=14cm]{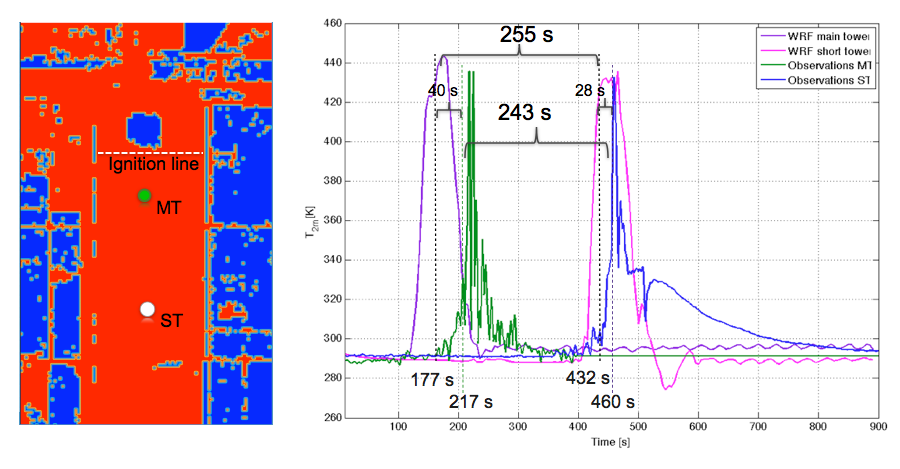}
\caption{Simulation of the FireFlux experiment \citep{Clements-2007-ODW} by
WRF-Fire. Left: map of landuse category for the experimental plot, with the
ignition line and the observation towers marked. Right: simulated and
measured temperature profiles at the location of the observation towers. The
simulated fire propagation takes 243\,s from tower MT to tower ST, while the
measured time is 255\,s (4.7\% difference). From \citet{Kochanski-2010-EFP}.
\label{fig:fireflux1}}
\end{figure}

\begin{figure}
\vspace*{2mm} \center\includegraphics[width=14cm]{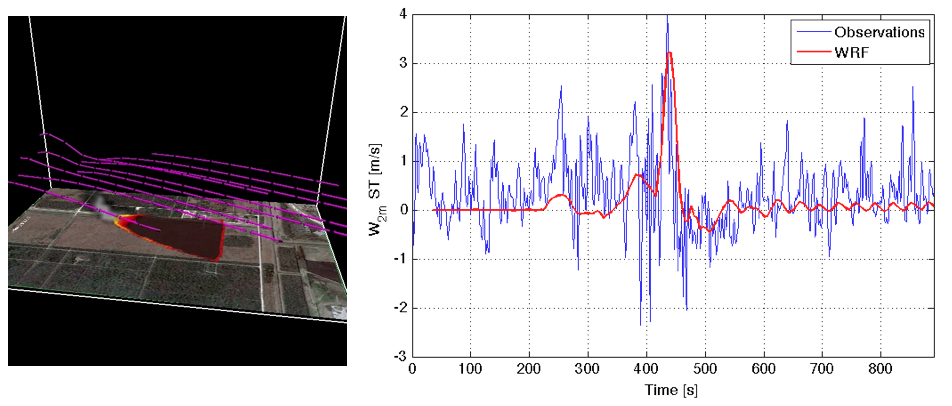}
\caption{Simulation of the FireFlux experiment \citep{Clements-2007-ODW} by
WRF-Fire. Left: surface heat flux and selected flowlines. Visualization in
VAPOR by Bed\v{r}ich Soused\'{\i}k. Surface image from Google Earth. Right:
vertical velocity at 2\,m height at tower ST. (See Fig.~\ref{fig:fireflux1}
left for location.) The simulation shows a good agreement with the
experiment. From \citet{Kochanski-2010-EFP}. \label{fig:fireflux2}}
\end{figure}

\begin{figure}
\vspace*{2mm} \center\includegraphics[width=10cm]{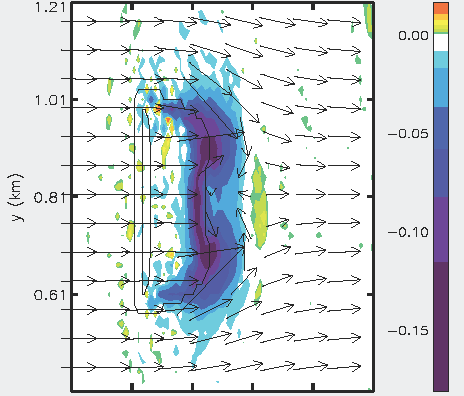}
\caption{X-Y section of wind vector at 18\,m and pressure perturbation 240\,s
after line ignition, initialized with uniform wind profile. The fire develops
two fingers due to wind direction inversion in the middle. From
\cite{Jenkins-2010-FDF}. \label{fig:jenkins1}}
\end{figure}

\begin{figure}
\vspace*{2mm} \center\includegraphics[width=14cm]{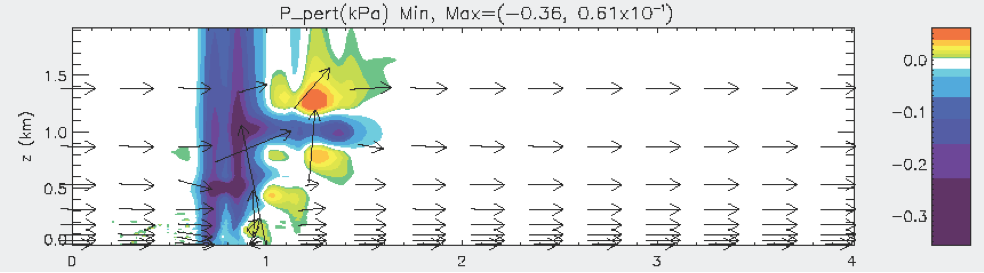}
\caption{X-Z section of wind vector and pressure perturbation at the
centerline for the fire in Fig.~\ref{fig:jenkins1}. From
\citet{Jenkins-2010-FDF}. \label{fig:jenkins2}}
\end{figure}

\begin{figure}
\vspace*{2mm} \center\includegraphics[width=10cm]{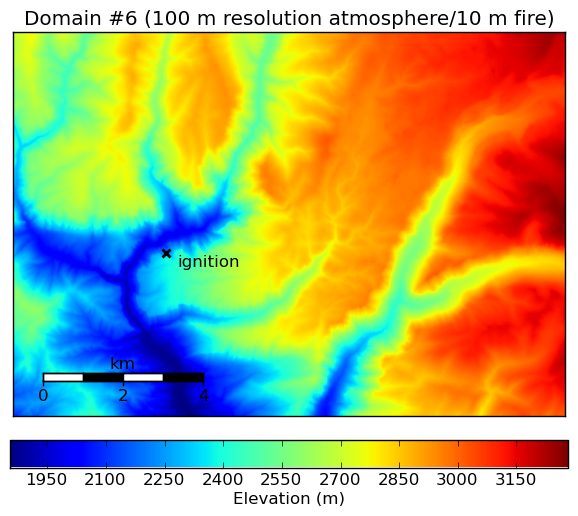}
\caption{Topography of the finest domain of the Meadow Creek fire simulation.
From \citet{Beezley-2010-SMC}. \label{fig:mc_domain6}}
\end{figure}

\begin{figure}
\vspace*{2mm} \center\includegraphics[width=12cm]{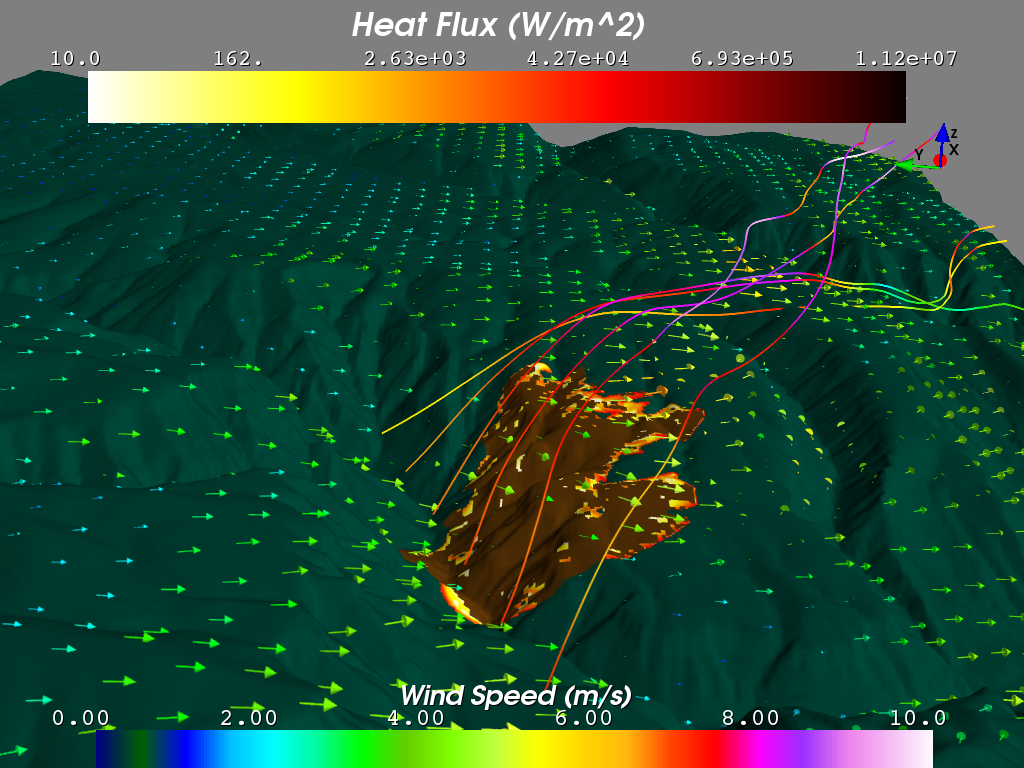}
\caption{The finest domain in the Meadow Creek fire simulation 5\,h after
ignition. Unburned fuel is displayed as green, burned fuel as brown. The heat
flux from the fire appears near the fire line. Arrows indicate the surface
winds, while streamlines show the atmospheric winds flowing over the fire
region. Visualization in MayaVi. From \cite{Beezley-2010-SMC}.
\label{fig:3dscene}}
\end{figure}

\begin{figure}
\vspace*{2mm} \center\includegraphics[width=12cm]{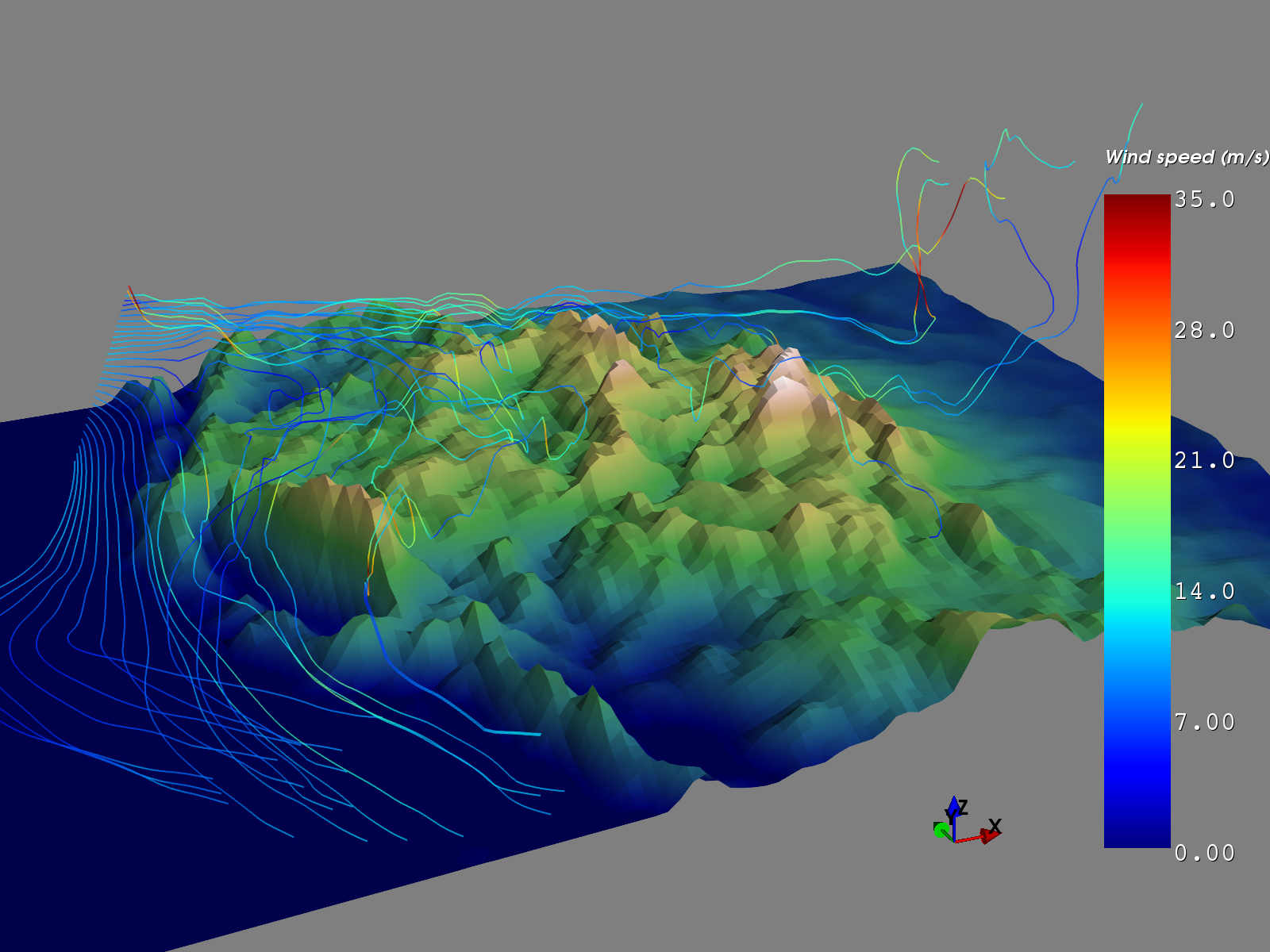}
\caption{The top level domain in the Meadow Creek fire simulation 5\,h after
ignition. Streamlines show the winds blowing East, over the Rocky Mountains
and South down the coast of California. Visualization in MayaVi. From
\citet{Beezley-2010-SMC}. \label{fig:3d_domain1}}
\end{figure}

\citet{Jenkins-2010-FDF} have demonstrated fireline fingering behavior for a
sufficiently long fireline (Figs.~\ref{fig:jenkins1}, \ref{fig:jenkins2}) on
an ideal example, with similar results as in
\citet{Clark-1996-CAF,Clark-1996-CAM}. \citet{Kochanski-2010-EFP} have
demonstrated the validity of \mbox{WRF-Fire} on a simulation of the
\citet{Clements-2007-ODW} FireFlux grass fire experiment and obtained good
agreement with data (Figs.~\ref{fig:fireflux1}, \ref{fig:fireflux2}).
\citet{Dobrinkova-2010-WAB} simulated a fire in Bulgarian mountains using
real meteorological and geographical data, and ideal fuel data.
\citet{Beezley-2010-SMC} simulated the 2010 Meadow Creek fire in Colorado
mountains using real data from online sources. Topography
(Fig.~\ref{fig:mc_domain6}) at up to 3\,m horizontal resolution was obtained
from the National Elevation Dataset (NED, \url{http://ned.usgs.gov}) and fire
fuel datasets from Landfire (\url{http://landfire.cr.usgs.gov}) at up to
10\,m resolution. Six nested domains were required to scale the simulation
down from the atmospheric initialization (32\,km) to the fire grid resolution
(10\,m). Cloud physics was enabled in domains 1--3. The fire subgrid
refinement ratio was 10\,times on the finest domain to capture fire surface
variables and for a gradual release of the heat flux near the fireline.
Realistic fire and atmosphere behavior was obtained (Figs.~\ref{fig:3dscene},
\ref{fig:3d_domain1}).\vspace*{-3.5mm}

\section{Discussion \label{sec:discussion}}\vspace*{-.5mm}

\subsection{Additional features}\vspace*{-.5mm}

WRF-Fire does not yet support canopy fire, although canopy fire colocated
with ground fire is contained in CAWFE. The reason was the desire to keep the
code as simple as possible early on and add features only as they can be
verified and validated. The support for canopy fire will be added in future.
Adding smoke from the fire to WRF is also under consideration. A list of
desired features and a record of the progress of the development are
maintained at \url{http://www.openwfm.org/wiki/WRF-Fire_development_notes}.
\vspace*{-.5mm}

\subsection{Atmosphere}\vspace*{-.5mm}

Rothermel's spread model (\ref{eq:spread}) assumes wind as if the fire was
not there. In practice, the wind was measured away from the fire. In a
coupled model, however, the feedback on the fire is from the wind that is
influenced by the fire. \citet{Clark-2004-DCA} noted that the horizontal wind
right above the fireline may even be zero, and proposed to take the wind from
a specified distance behind the fireline. Also, the strong heat flux from
fire disturbs the logarithmic wind profile, and the rate of spread as a
function of wind at a specific altitude may not be a good approximation;
rather, the fire spread may depend more strongly on the complete wind profile
\citep{Jenkins-2010-FDF} and on turbulence \citep{Sun-2009-IFC}. The
assumption of horizontal homogeneity in the Monin-Obukhov similarity theory
is not satisfied here; the horizontal dimension of the active part of fire is
not orders of magnitude larger than the boundary layer height as required,
and it may be in fact smaller. Another indication that the Monin-Obukhov
theory may not apply for fires is a strong drop in the heat transfer in the
case of strong temperature gradients, shown in our preliminary tests.

Horizontal wind could be interpolated vertically to different heights for different
fuels like in CAWFE model, which takes the wind from different mesh levels
for different fuels. However, here we follow a classical approach of
\citet{Rothermel-1972-MMP} and \citet{Baughman-1980-EMW},
where the wind speed is evaluated at the common 6.1m height, and then
converted to the mid-flame height using the fuel-specific wind correction factors.

Very strong vertical components of the wind caused by the fire result in the
need for short time steps to avoid violation of the vertical CFL condition
(Sect.~\ref{sec:timestep}). It would be interesting to couple the fire module
also with the Non-hydrostatic Mesoscale Model (NMM) core of WRF, which is
implicit in the vertical direction \citep{Janjic-2005-NWN}, and it may
perform better in the presence of strong convection \citep{Litta-2008-SST}.
The ARW core is semi-implicit in the vertical direction in the
vertical wind component and the geopotential.

\subsection{Fire}

The more recent \citet{Scott-2005-SFB} fuel categories are more detailed than
\citet{Anderson-1982-ADF} categories, they are supported by BehavePlus, and
fuel maps using them are available from Landfire. But instead of describing
additional categories in \texttt{namelist.fire}, it may be more useful to
support the import of fuel files from BehavePlus, which is also well suited
for editing and diagnosing fuel models. More accurate fuel models
\citep{Albini-1995-CLF,Clark-1996-CAF}, including those in BehavePlus,
consider fuels to be mixtures of components with different burn times,
which results in a different heat release curve.

While the spread rate of established fire in the simulation of the FireFlux
experiment was reasonably close, the simulated fire still arrived at the
observation towers too soon \citep{Kochanski-2010-EFP}, because it started
too quickly. A better parametrization of the ignition process seems to be in
order. The fire spread in the Meadow Creek fire simulation was also too fast,
but for a different reason. It is well known that the actual spread rates of
wildland fires tend to be lower than the spread rates in simulations, which
are derived from laboratory experiments. This effect might be attributed to
irregularities on scales not captured by the simulation
\cite[p.~34]{Finney-1998-FFA}, including granularity of the fuel supply not
reflected in the data. Refining the semi-empirical model from detailed
numerical simulations and parametrizing complex fire behavior are suggested
important research areas.

The computation of the heat fluxes in (\ref{eq:sensible-heat-flux}) and
(\ref{eq:latent-heat-flux}) does not take into account the evaporation of
moisture present in the fuel, only the production of water by burning of
hydrocarbons. This error is typically just few \%, however, which is small in
comparison with other uncertainties. The fuel models should be dynamic (with
variable fuel moisture) as in BehavePlus. \citet{Coen-2005-SBE} added an
explicit diurnal cycle for the moisture into CAWFE. Here, moisture content
could be coupled with existing WRF land surface models, which could take into
account air humidity and precipitation. The radiative and convective parts of
the sensible heat flux should be treated differently. The release of surface
heat and moisture into the atmosphere are already present in WRF soil models.
Their scale, however, is different from the powerful heat release from a
fire.

\subsection{Numerical methods}

In a numerical implementation, the level-set method is global, unlike
tracers, which move locally. In spite of the fact that the level-set equation
determines the fire spread locally from the spread rate at the fireline, the
behavior of the fireline depends slightly on the wind, the fuel, and the
level set function in certain other locations from previous time steps,
because of the discretization errors and the artificial diffusion. This
nonlocal behavior has not been practically significant, however.

The fuel fraction calculation (\ref{eq:approx-fraction}) can have significant
error in the fire subgrid cells near the fireline, which will to some degree
average out over the atmospheric mesh cells. Rigorous error analysis will be
done elsewhere. We are currently testing an alternative method which is
always second order in the sense that it is exact when the time from ignition
and the level-set function are linear in space. The alternative method is
more computationally expensive, but, on the other hand, it might allow to decrease the
subgrid refinement ratio; with large meshes, it is possible to run against
32\,bit integer limits.

\subsection{Data assimilation}

Data assimilation for wildland fires is an area of great interest.
Methodologies for a reaction-diffusion model were proposed based on the
ensemble Kalman filter (EnKF) and the particle filter
\citep{Mandel-2004-NDD}. Unfortunately, statistical perturbations can cause
spurious fires, which do not dissipate. Combination of the EnKF with Tikhonov
regularization alleviates the problem somewhat
\citep{Johns-2008-TEK,Mandel-2009-DAW}, but the resulting method is still not
robust enough. A new method, called morphing EnKF and based on combined
amplitude and displacement correction \citep{Beezley-2008-MEK}, was shown to
work with \mbox{WRF-Fire} \citep{Mandel-2009-DAW}, and it is under continued
development \citep{Mandel-2010-FFT,Mandel-2011-SME}. We are not aware of any
work elsewhere on data assimilation for a coupled fire-atmosphere model.
Particle filters were proposed for discrete cell-based fire models
\citep{Bianchini-2006-IPM,Gu-2009-SEU}, using fitness functions involving the
area burned rather than intensities of physical variables.

Starting the model from a known fire perimeter is important for many
potential users. This can be understood as a data assimilation problem, but
we are considering a simpler method for this particular case: prescribe the
fire history up to the time of the given perimeter to allow the atmospheric
conditions to evolve, then allow the coupled model take over. Tools to produce
such artificial fire history are being developed. Possibly the simplest
alternative is an interpolation from a given ignition point and time to the
given perimeter. A more complex version would run the fire model (without
atmosphere) backwards in time and attempt to find the ignition point
automatically. The latter approach could be also interesting for forensic
purposes.

\conclusions We have described the coupled atmosphere-fire model
\mbox{WRF-Fire}. The software is publicly available and it supports both
ideal and real runs. Visualization and diagnostic utilities are available.
Currently, the model is suitable for research and education purposes.
Validation is in progress.

\begin{acknowledgements}
The authors would like to thank John Michalakes for developing the support for the
refined surface fire grid in WRF and information about WRF algorithms,
Ned~Patton for providing a copy of
his prototype code, and Janice~Coen for providing a copy of CAWFE, liason
with NCAR, and useful suggestions. Other contributions to the model are
acknowledged by bibliographic citations in the text. We would like to thank also
Mary Ann Jenkins for reading this paper and suggesting improvements.  This research was
supported by NSF grant AGS-0835579 and NIST Fire Research Grants Program
grant 60NANB7D6144.

\end{acknowledgements}

\bibliographystyle{copernicus}
\bibliography{../../references/geo,../../references/other}


\clearpage

\clearpage

\begin{table}[t]
\caption{Fuel properties. The notation is from \citet{Rothermel-1972-MMP}
except as indicated. The identifiers are as used in \mbox{WRF-Fire} and
CAWFE. In the input files, some quantities are given in English units per
\citet{Rothermel-1972-MMP}; see \citet[p.~A-5]{Wang-2010-AUG}.
\label{tab:fuels}} \vskip4mm
\begin{center}
\begin{tabular}{lll}
\tophline
symbol & description & identifier         \\
\middlehline
$a$            & wind adjustment factor \citep{Baughman-1980-EMW} & \texttt{windrf}       \\
               & from 6.1\,m to midflame length                   &                       \\
$w$            & fuel weight (i.e., burn time) (s)                &                       \\
               & 40\% decrease of fuel in 10\,min for $w=1000$    & \texttt{weight}       \\
$w_{\ell}$     & total fuel load (kg\,m$^{-2}$)                   & \texttt{fgi}          \\
$\delta_{m}$   & fuel depth (m)                                   & \texttt{fueldepthm}   \\
$\sigma$       & fuel particle surface-area-to-volume ratio (1/m) & \texttt{savr}         \\
$M_{x}$        & moisture content of extinction (1)               & \texttt{fuelmce}      \\
$\rho_{\rm P}$ & ovendry fuel particle density (kg\,m$^{-3}$)     & \texttt{fueldens}     \\
$S_{\rm T}$    & fuel particle total mineral content (1)          & \texttt{st}           \\
$S_{\rm E}$    & fuel particle effective mineral content (1)      & \texttt{se}           \\
$h$            & fuel heat contents of dry fuel (J\,kg$^{-1}$)    & \texttt{cmbcnst}      \\
$M_{\rm f}$    & fuel particle moisture content (1)               & \texttt{fuelmc\_g}    \\
\bottomhline
\end{tabular}
\end{center}
\end{table}

\clearpage

\begin{table}[t]
\caption{\scriptsize Computation of the fire spread rate factors in (\ref{eq:spread})
from the fuel properties (Table~\ref{tab:fuels}), the wind speed $U$ at
6.1\,m, and the terrain slope $\tan\phi$. All equations are from
\citet{Rothermel-1972-MMP} unless otherwise indicated. All input quantities
are first converted from metric to English units (BTU-lb-ft-min) to avoid
changing the numerous constants in the \citet{Rothermel-1972-MMP}
computations. Further, following CAWFE, the wind is limited to between 0 and
30\,m\,s$^{-1}$ and the slope is limited to nonnegative values.
\label{tab:spread}}
\begin{center}\scalebox{.88}[.88]{
\begin{tabular}{lll}
\tophline
equation & description & source\\
\middlehline
$R_0=\frac{I_R \xi}{\rho_b \varepsilon Q_{ig}}$                                                                               & \textbf{spread rate without wind} & Eq.~(52)                    \\[1.5mm]
$\xi = \frac{\exp\left[\,  \left(  0.792+0.681\sigma^{0.5}\right)  \left( \beta+0.1\right)\,  \right] }{192+0.2595\sigma}$    & propagating flux ratio            & Eq.~(42)                    \\[1.5mm]
$I_R = \Gamma w_nh\eta_M \eta_s$                                                                                              & reaction intensity                & Eq.~(52)                    \\[1.5mm]
$\eta_s = 0.174S_{e}^{-0.19}$                                                                                                 & mineral damping coefficient       & Eq.~(30)                    \\[1.5mm]
$\eta_M = 1-2.59\frac{M_f}{M_x}+5.11\left(  \frac{M_f}{M_x}\right)^2-3.52\left(  \frac{M_f}{M_x}\right)^3$                    & moisture damping coefficient      & Eq.~(29)                    \\[1.5mm]
$w_n = \frac{w_0}{1+S_T}$                                                                                                     & fuel loading net of minerals      & Eq.~(24)                    \\[1.5mm]
$w_0 = \frac{w_{\ell}}{1+M_f}$                                                                                                & total fuel load net of moisture   & from CAWFE                  \\[1.5mm]
$\Gamma=\Gamma_{\max}\left(  \frac{\beta}{\beta_{op}}\right)^A \exp\left[A\left(  1-\frac{\beta}{\beta_{op}}\right)  \right]$ & optimum reaction velocity         & Eq.~(36)                    \\[1.5mm]
$\Gamma_{\max}=\frac{\sigma^{1.5}}{495+0.594\sigma^{1.5}}$                                                                    & maximum reaction velocity,        & Eq.~(36)                    \\[1.5mm]
$\beta = \frac{\rho_b}{\rho_P}$                                                                                               & packing ratio                     & Eq.~(31)                    \\[1.5mm]
$\rho_b = \frac{w_0}{\delta}$                                                                                                 & oven dry bulk density             & Eq.~(40)                    \\[1.5mm]
$A=\frac{1}{4.77\sigma^{0.1}-7.27}$                                                                                           &                                   & Eq.~(39)                    \\[1.5mm]
$\varepsilon=\exp\left(  -\frac{138}{\sigma}\right) $                                                                         & effective heating number          & Eq.~(14)                    \\[1.5mm]
$Q_{ig}=250\beta+1116M_f$                                                                                                     & heat of preignition               & Eq.~(12)                    \\[1.5mm]
$\phi_W = C\max U_a^{\beta}\left(  \frac{\beta}{\beta_{op}}\right)^E$                                                         & \textbf{wind factor}              & Eq.~(47)                    \\[1.5mm]
$C=7.47\exp\left(  -0.133\sigma^{0.55}\right)$                                                                                &                                   & Eq.~(48)                    \\[1.5mm]
$U_a = aU$                                                                                                                    & adjustment to midflame height     & Table~\ref{tab:fuels} here  \\[1.5mm]
$E = 0.715\exp\left(  -3.59\times 10^{-4}\sigma\right)$                                                                       &                                   & Eq.~(50)                    \\[1.5mm]
$\phi_S = 5.275\beta^{-0.3}\tan^{2}\phi$                                                                                          & \textbf{slope factor}             & Eq.~(511q)                    \\
\bottomhline
\end{tabular}}
\end{center}
\end{table}

\clearpage

\clearpage

\clearpage

\clearpage

\clearpage

\clearpage

\clearpage

\clearpage

\clearpage

\clearpage

\clearpage

\clearpage

\end{document}